\documentclass[aps,prl,twocolumn,groupedaddress,showpacs,floatfix]{revtex4}
\usepackage{graphicx}
\usepackage{color}
\usepackage{amssymb}

\begin{document}
%\preprint{}
\title{Nuclear transparency and effective kaon-nucleon cross section from the $A(e,e^\prime K^{+})$ reaction}
\author{Nuruzzaman$^{1}$}
\author{D.~Dutta$^{1}$}
\author{J.~Arrington$^{2}$}
\author{R.~Asaturyan$^{3}$}
\author{F.~Benmokhtar$^{4}$}
\author{W.~Boeglin$^{5}$}
\author{P.~Bosted$^{6}$}
\author{A.~Bruell$^{6}$}
\author{B.~Clasie$^{7}$}
\author{M.~E.~Christy$^{8}$}
\author{E.~Chudakov$^{6}$}
\author{M.~M.~Dalton$^{9}$}
\author{A.~Daniel$^{10}$}
\author{D.~Day$^{11}$}
\author{L.~El~Fassi$^{2}$}
\author{R.~Ent$^{6}$}
\author{H.~C.~Fenker$^{6}$}
\author{J.~Ferrer$^{12}$}
\author{N.~Fomin$^{11}$}
\author{H.~Gao$^{7,13}$}
\author{K.~Garrow$^{14}$}
\author{D.~Gaskell$^{6}$}
\author{C.~Gray$^{9}$}
\author{T.~Horn$^{4,6}$}
\author{G.~M.~Huber$^{15}$}
\author{M.~K.~Jones$^{6}$}
\author{N.~Kalantarians$^{10}$}
\author{C.~E.~Keppel$^{6,8}$}
\author{K.~Kramer$^{13}$}
\author{Y.~Li$^{10}$}
\author{Y.~Liang$^{16}$}  
\author{A.~F.~Lung$^{6}$}
\author{S.~Malace$^{8}$}
\author{P.~Markowitz$^{5}$}
\author{A.~Matsumura$^{17}$}
\author{D.~G.~Meekins$^{6}$}
\author{T.~Mertens$^{18}$}
\author{T.~Miyoshi$^{10}$}
\author{H.~Mkrtchyan$^{3}$}
\author{R.~Monson$^{19}$}
\author{T.~Navasardyan$^{3}$}
\author{G.~Niculescu$^{12}$}
\author{I.~Niculescu$^{12}$}
\author{Y.~Okayasu$^{17}$}
\author{A.~K.~Opper$^{16}$}
\author{C.~Perdrisat$^{20}$}
\author{V.~Punjabi$^{21}$}
\author{X.~Qian$^{13}$}
\author{A.~W.~Rauf$^{22}$}
\author{V.~M.~Rodriquez$^{10}$}
\author{D.~Rohe$^{18}$}
\author{J.~Seely$^{7}$}
\author{E.~Segbefia$^{8}$}
\author{G.~R.~Smith$^{8}$}
\author{M.~Sumihama$^{17}$}
\author{V.~Tadevosyan$^{3}$}
\author{L.~Tang$^{6,8}$}
\author{V.~Tvaskis$^{6,8}$}
\author{W.~F.~Vulcan$^{6}$}
\author{F.~R.~Wesselmann$^{21}$}
\author{S.~A.~Wood$^{6}$}
\author{L.~Yuan$^{8}$}
\author{X.~C.~Zheng$^{2}$}

\affiliation{$^{1}$Mississippi State University, Mississippi State, MS, USA}
\affiliation{$^{2}$Argonne National Laboratory, Argonne, IL, USA}
\affiliation{$^{3}$Yerevan Physics Institute, Yerevan, Armenia}
\affiliation{$^{4}$University of Maryland, College Park, MD, USA}
\affiliation{$^{5}$Florida International University, Miami, FL, USA}
\affiliation{$^{6}$Thomas Jefferson National Laboratory, Newport News, VA, USA}
\affiliation{$^{7}$Laboratory for Nuclear Science, Massachusetts Institute of Technology, Cambridge, MA, USA}
\affiliation{$^{8}$Hampton University, Hampton, VA, USA}
\affiliation{$^{9}$University of the Witwatersrand, Johannesburg, South Africa}
\affiliation{$^{10}$University of Houston, Houston, TX, USA}
\affiliation{$^{11}$University of Virginia, Charlottesville, VA, USA}
\affiliation{$^{12}$James Madison University, Harrisonburg, VA, USA}
\affiliation{$^{13}$Triangle Universities Nuclear Laboratory, Duke University, Durham, NC, USA}
\affiliation{$^{14}$TRIUMF, Vancouver, British Columbia, Canada}
\affiliation{$^{15}$University of Regina, Regina, Saskatchewan, Canada}
\affiliation{$^{16}$Ohio University, Athens, OH, USA}
\affiliation{$^{17}$Tohoku University, Sendai, Japan}
\affiliation{$^{18}$Basel University, Basel, Switzerland}
\affiliation{$^{19}$Central Michigan University, Mount Pleasant, MI, USA}
\affiliation{$^{20}$College of William and Mary, Williamsburg, VA, USA}
\affiliation{$^{21}$Norfolk State University, Norfolk, VA, USA}
\affiliation{$^{22}$University of Manitoba, Winnipeg, Manitoba, Canada}

%%%%%%%%%%%%%%%%%%%%%%%%%%%%%%%%%%%%%%%%%%%%%%%%%%%%%%%%%%%%%%%%%%%%%%%%%%%%%%%%%%%%%%%%%%%%%%%%%%%%%%%%
%%%%%%%%%%%%%%%%%%%%%%%%%%%%%%%%%%%%%%%%%%%%%%Abstract%%%%%%%%%%%%%%%%%%%%%%%%%%%%%%%%%%%%%%%%%%%%%%%%%%
%%%%%%%%%%%%%%%%%%%%%%%%%%%%%%%%%%%%%%%%%%%%%%%%%%%%%%%%%%%%%%%%%%%%%%%%%%%%%%%%%%%%%%%%%%%%%%%%%%%%%%%%
\begin{abstract}
We have determined the transparency of the nuclear medium to kaons from $A(e,e^{'} K^{+})$ measurements on $^{12}$C, $^{63}$Cu,
and $^{197}$Au targets. The measurements were performed at the Jefferson Laboratory and span a range in four-momentum-transfer squared
Q$^2$=1.1 -- 3.0 GeV$^2$. The nuclear transparency was defined as the ratio of measured kaon electroproduction cross sections with
respect to deuterium, ($\sigma^{A}/\sigma^{D}$). We further extracted the atomic number ($A$) dependence of the transparency
as parametrized by $T= (A/2)^{\alpha-1}$ and, within a simple model assumption, the in-medium effective kaon-nucleon cross sections.
The effective cross sections extracted from the electroproduction data are found to be smaller than the free cross sections
determined from kaon-nucleon scattering experiments, and the parameter $\alpha$ was found to be significantly larger than those
obtained from kaon-nucleus scattering. We have included similar comparisons between pion- and proton-nucleon effective cross sections
as determined from electron scattering experiments, and pion-nucleus and proton-nucleus scattering data.
\end{abstract}

%\keywords{}

\maketitle

The propagation of hadrons in the nuclear medium is an essential element to the understanding of the nuclear many-body system as constructed
from the more basic meson-nucleon and nucleon-nucleon amplitudes. Quasi-free electron scattering from nuclei provides an excellent tool
for a microscopic examination of such hadron propagation effects in the nuclear medium. The combination of the relative weakness of the
electromagnetic probe and a well-understood interaction at the production vertex are well-known advantages of the electron scattering
process. Due to this, quasi-free production can be viewed as tagging a source of hadrons emerging from throughout the nuclear volume,
with minimal disruption of the nuclear system. Thus, over the past two decades, proton propagation in nuclei has been studied extensively
using quasi-free electron scattering~\cite{bates,ne18,jlabp1,jlabp2}. More recently, pion propagation in nuclei was studied using exclusive
electroproduction of pions from nuclei~\cite{jlabpi1,jlabpi2}. 

Electroproduction of positively-charged kaons, $K^+$, from nuclei can provide an additional, strangeness, degree of freedom, which is
inaccessible with nucleon and/or pion knockout. It is well known that strangeness production provides an unique window on the nuclear many-body
problem via access to energy levels that protons and neutrons cannot occupy. Moreover, the $K^+$--nucleon ($K^+-N$) interaction is itself also relatively weak and varies smoothly with energy~\cite{dover77}, which simplifies the extraction of the in-medium $K^+$-nucleus
interaction, as well as the physics of hadron formation. Typically, the $K^+$--nucleus interaction has been studied using $K^+$ scattering
from nuclear targets, and these experiments can be considered as analogous to electron scattering since both involve weakly interacting probes.

In spite of the $K^+-N$ interaction being relatively weak and relatively free of resonance structure, it has proven difficult to obtain a successful
description of kaon-nucleus scattering from the elementary $K^+-N$ interaction~\cite{skg1}. Both differential and total cross section
measurements~\cite{kccexpt1,kccexpt2,kccexpt3,kccexpt4,kccexpt5,kccexpt6} from  $K^+$ scattering experiments show significant discrepancies
when they are compared to theoretical calculations~\cite{skg1,kccth1,kccth2,kccth3}. Even when taking the ratio of total cross sections for
nuclear and deuterium targets~\cite{kccexpt1,kccexpt3,kccexpt4,kccexpt5,kccexpt6}, where some of the theoretical uncertainties
are expected to cancel, experimental results are found to be systematically larger than the theoretical expectations~\cite{skg1, kccth1,kccth2,kccth3}.
Given the apparent failure of conventional nuclear physics models to account for the data, several exotic mechanisms have been proposed by
various authors~\cite{ernst95}. These include modification of the nucleon size in the nuclear medium~\cite{skg1}, reduced meson masses in
the nuclear medium~\cite{kccth4}, meson exchange currents~\cite{kccth5,kccth6}, long range correlations~\cite{kccth7}, and various other
mechanisms~\cite{kccth8}.   

The availability of kaon production data in heavy-ion collisions~\cite{sisdata} renewed the interest in the properties of kaons in nuclear
matter, since it was argued that such data have the capacity to signal chiral symmetry restoration or give information on the possibility
of kaon condensation in neutron stars~\cite{kaoncond}. Similar to what was found for the $K^+$ scattering experiments, the comparison between
the heavy-ion collision data and calculations~\cite{sisth} indicates that the calculations substantially underestimate the experimental data.
Recently, a calculation in the framework of the quark-meson coupling model~\cite{sibirtsev}, going beyond the conventional meson-nucleon
nuclear many-body framework, was able to describe these kaon production data in heavy-ion collisions.

The electroproduction of kaons from nuclei is an excellent alternative to both kaon-nucleus scattering and kaon production in heavy-ion collisions,
particular in order to explore the propagation of kaons through the nuclear medium. It may help determine if the noted discrepancies between theory
and experiment are related to the details of the reaction mechanism or if they are due to the various approximations made in calculating the cross sections. The electroproduction of kaons from light nuclei~\cite{eek1,eek2} and $^{12}$C~\cite{hinton} at Jefferson Lab (JLab) has been reported before. In this article, we report the results from the quasi-free electroproduction of kaons from heavy nuclei. We report the atomic mass number, $A$, and four-momentum transfer squared, $Q^2$, dependence of the nuclear transparency of kaons as extracted from the $A(e,e'K^+)$ process. 
%The measurements were performed on the elementary proton and deuteron ($^1$H and $^2$H) targets, as well as on heavier nuclei $^{12}$C, $^{27}$Al, $^{63}$Cu and $^{197}$Au. The data map a $Q^2$ range of 1.1 $\mathrm{to}$ 3.0 $\mathrm{GeV}^2$. We have defined the nuclear transparency as the the ratio of the heavy-nucleus cross section to the deuteron cross section to reduce the impact of non-isoscalar effects, but we will also list results with respect to the free proton target. The $A$ dependence of the data is then further described from the simple ansatz $T = (A/2)^{\alpha -1}$, where $\alpha$ is used to parametrize the nuclear cross section. This simple ansatz is equivalent to the more commonly used $\sigma_N = \sigma_0A^{\alpha}$ parameterization, with $\sigma_N$ the nuclear cross section and $\sigma_{0}$ the elementary (isoscalar) nucleon cross section~\cite{carroll}.

%%%%%%%%%%%%%%%%%%%%%%%%%%%%%%%%%%%%%%%%%%%%%%%%%%%%%%%%%%%%%%%%%%%%%%%%%%%%%%%%%%%%%%%%%%%%%%%%%%%%%%%%
%%%%%%%%%%%%%%%%%%%%%%%%%Description of the experiment%%%%%%%%%%%%%%%%%%%%%%%%%%%%%%%%%%%%%%%%%%%%%%%%%%
%%%%%%%%%%%%%%%%%%%%%%%%%%%%%%%%%%%%%%%%%%%%%%%%%%%%%%%%%%%%%%%%%%%%%%%%%%%%%%%%%%%%%%%%%%%%%%%%%%%%%%%%
The experiment was carried out in Hall C at JLab~\cite{JLab} in 2004. The kinematics are shown in Table~\ref{table1}. The experiment was
designed to measure the nuclear transparency of pions following the $A(e,e^\prime \pi^+$) reaction, but additional particle identification allowed
us to simultaneously measure electro-produced positively-charged kaons. The details about the experiment can be found in Ref.~\cite{jlabpi2}.
The scattered electrons were detected in the short orbit spectrometer (SOS) and the electroproduced pions and kaons were detected in the high momentum
spectrometer (HMS). A detailed  description of the spectrometers and the spectrometer hardware is given in~\cite{arr98,dutta03}. The detector package of
the HMS was equipped with a gas \v{C}erenkov, and an aerogel \v{C}erenkov~\cite{aero} and a lead glass calorimeter, for $p/K^+$/$\pi^+$ separation.
For this experiment, the gas \v{C}erenkov was filled with $C_4F_{10}$ gas at 97~kPa. The corresponding index of refraction is 1.00137, resulting in momentum
thresholds of 2.65 GeV for $\pi^+$ and 9.4 GeV for $K^+$. The aerogel \v{C}erenkov material had an index of refraction of 1.015, giving thresholds of
0.8 GeV for pions and 2.85 GeV for kaons. It was used to distinguish between $\pi^+$ and $K^+$ for central momentum settings below 3.1 GeV and
to distinguish $\pi^+$ and $K^+$ from protons for momentum settings above 3.1 GeV. For momentum settings above 3.1 GeV, the gas \v{C}erenkov counter
was used to distinguish between kaons and pions. Coincidence timing between the HMS and SOS, using multiple scintillator hodoscopes, could be used to
distinguish between $K^+$ and protons with momenta up to $\approx$ 3 GeV.

\begin{table}
\caption{The central kinematics of the experiment. $E_e$ ($E_{e^\prime}$) is the incident (scattered) electron energy,
$\theta_{e^\prime}^{\rm SOS}$ ($\theta_{HMS}$) is the scattered electron angle (kaon angle), $p_{K^+}$ is the (central) kaon momentum, and $|t|$ is the Mandelstam four-momentum transfer. $\theta_{K^+}$ is the central kaon angle expected from the electron kinematics.}
%\begin{ruledtabular}
\begin{tabular}{|c|c|c|c|c|c|c|c|}\hline
 $Q^2$ & $-t$ & $E_e$ & $\theta_{e^\prime}^{\rm SOS}$ &$E_{e^\prime}$ &
 $\theta_{\rm HMS}$ & $p_{K^+}$ & $\theta_{K^+}$\\
 GeV$^2$ & GeV$^2$ & GeV & deg & GeV & deg & GeV & deg  \\\hline
1.1 &0.05 &4.021 &27.76 &1.190 &10.61 &2.793 &10.58\\
2.1 &0.16 &5.012 &28.85 &1.730 &13.44 &3.187 &13.44\\
3.0 &0.29 &5.012 &37.77 &1.430 &12.74 &3.418 &12.74\\\hline
\end{tabular}
%\end{ruledtabular}  
%\vspace{-0.5cm}
\label{table1} 
\end{table}

\begin{figure}[h]
	\centering
	\includegraphics[width=86mm]{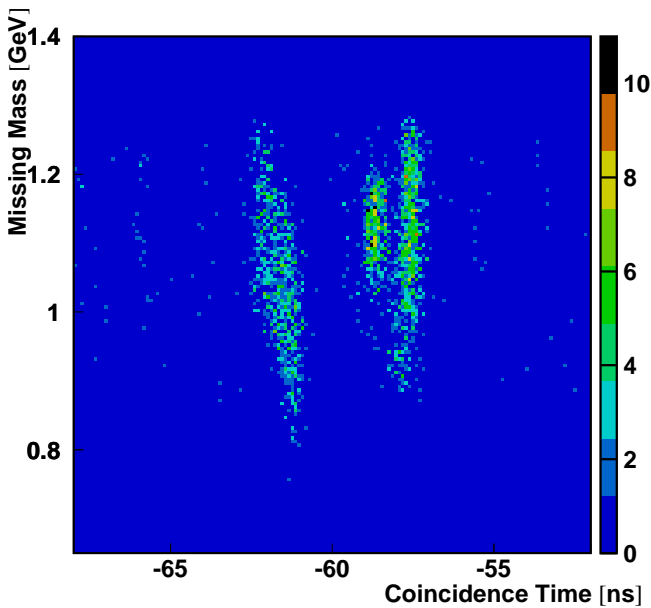}
	\vspace{0.43ex}
	\caption{(Color Online)The missing mass versus coincidence time for $^{12}C(e,e'K^+)$ events at $Q^2=$ 2.1 GeV$^2$.}
	\label{fig:pid}
\end{figure}

The SOS gas \v{C}erenkov counter was used to select the scattered electrons with an efficiency of better than 99.2\%. The charged kaons were selected using
the HMS  aerogel~\cite{aero} and gas \v{C}erenkov counters, the net kaon detection efficiency was determined from the product of the HMS tracking efficiency, the aerogel, and gas \v{C}erenkov efficiencies, and the time-of-flight cut efficiency. The kaon detection efficiency was found to be 82\% -- 85\%. The pion contamination in the kaon samples was estimated to be $<$~1\%. The spectrometer acceptance was determined with a relative uncertainty of 1\% between targets using a Monte Carlo simulation of the experimental apparatus, as described below. For each run,
the $e-K^+$ coincidence events were corrected for random coincidences, by subtracting their contribution as determined using out-of-time events. The charge weighted coincidence yield was also corrected for blocked coincidences
($<$0.7\%), loss of synchronization between detectors ($<$1.0\%), trigger inefficiency ($<$0.5\%), electronic dead time ($<$1.0\%),
computer dead time ($<$25\%, known to much better than 1\%), tracking inefficiency ($<$4.0\%) and particle absorption in the spectrometer materials
(5.0\%, known to better than 1.0\%). Further a cut on the missing-mass spectrum corresponding to the $\Lambda^0$ and $\Sigma^0/\Sigma^-$ hyperons was used to
select the kaon events. The uncertainty from variation of these cuts was 2.5\%. A typical two-dimensional spectrum of missing mass versus coincidence time (difference in time between the kaon reaching the HMS
and the electron reaching the SOS spectrometers) is shown in Fig~\ref{fig:pid}.
For $A>1$ targets we are unable to separate the $\Lambda^0$ and $\Sigma^0/\Sigma^-$ hyperons, consequently, we integrate over all three hyperons.
 %The events selected after applying all the cuts are shown by the red crosses.

%%%%%%%%%%%%%%%%%%%%%%%%%%%%%%%%%%%%%%%%%%%%%%%%%%%%%%%%%%%%%%%%%%%%%%%%%%%%%%%%%%%%%%%%%%%%%%%%%%%%%%%%
%%%%%%%%%%%%%%%%%%%%%%%%%%%%%%%%%%%%%%%%%%%%Simulation%%%%%%%%%%%%%%%%%%%%%%%%%%%%%%%%%%%%%%%%%%%%%%%%%%
%%%%%%%%%%%%%%%%%%%%%%%%%%%%%%%%%%%%%%%%%%%%%%%%%%%%%%%%%%%%%%%%%%%%%%%%%%%%%%%%%%%%%%%%%%%%%%%%%%%%%%%%
The standard Hall C Monte Carlo simulation code SIMC was used to simulate the experimental apparatus~\cite{ben_thesis}. The $^1H(e,e'K^+)$ cross section
required as model input was iterated until there was good agreement between simulation and the experimental data. The iteration was performed separately
for each of the kinematic ($Q^2$) settings in Table \ref{table1}. A parametrization of the $^1H(e,e'K^+)$ cross section from previous data~\cite{hinton,mohring} was used as the starting model. The model dependent uncertainty was estimated to be $\sim$ 4\%.

The Fermi motion of the nucleons in $A>1$ targets was simulated by folding the elementary cross section with a spectral function for the target. For each target, an appropriate Independent Particle Shell Model (IPSM) spectral function was used~\cite{dutta03}. The simulation includes several corrections such as kaon decay within the spectrometer, external and internal bremsstrahlung radiation. It also included corrections due to various reaction mechanism effects such as Coulomb distortions. More details on the Monte Carlo simulation can be found in Ref.~\cite{ben_thesis}.

\begin{figure}[h]
	\centering
	\includegraphics[width=86mm]{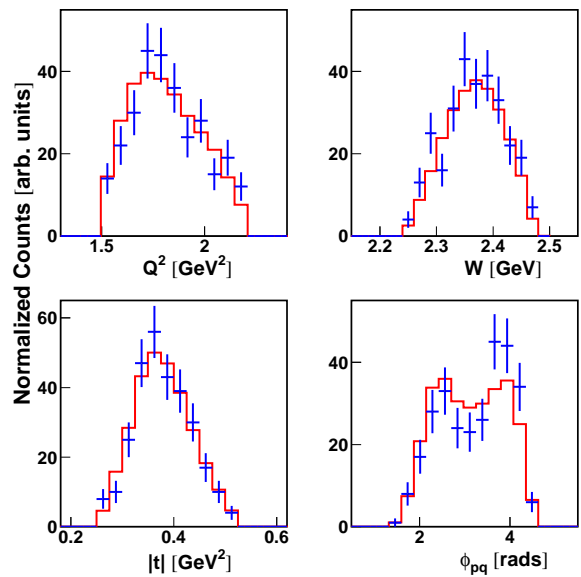}
	\vspace{0.43ex}
	\caption{(Color online) The $^{12}C(e,e'K^+)$ data and Monte Carlo simulations at $Q^2=$ 2.1 GeV$^2$ are compared as a function of $Q^2$, invariant-mass $W$, Mandelstam four-momentum transfer $|t|$, and the angle between momentum transfer vector and electro-produced kaon $\phi_{pq}$. Experimental data are shown as blue crosses, along with their statistical
uncertainty. Simulations are represented by the red histograms.}
	\label{fig:comp}
\end{figure}

The simulation was able to reproduce the shapes of the measured $Q^2$, $W$, $|t|$ and $\phi_{pq}$ distributions reasonably well for all targets and $Q^2$ settings. Here, $W$ is the invariant-mass, $|t|$ is the  Mandelstam four-momentum transfer and $\phi_{pq}$ is the angle between momentum transfer vector and electro-produced kaon. A typical comparison between data and simulation is shown in Fig.~\ref{fig:comp}, for the $^{12}$C nucleus at $Q^2$ = 2.1 GeV$^2$. The ratio of $\Lambda$ and $\Sigma$ hyperon production of kaons has been calculated by using Hydrogen data and the same ratio has been used for all other targets. In order to extract nuclear transparencies from the experimental yields, the cross section for the bound proton must be corrected for the effects of Fermi motion, the off-shell properties of the proton and the acceptances of the spectrometers. The nuclear transparency for a given target, with atomic number, $A$, was defined as:

\begin{equation} \label{equ:nucltransp}
T = 
\frac{{\left( \bar Y/ \bar Y_{\rm MC} \right)_A} }{
{\left( \bar Y/ \bar Y_{\rm MC} \right)_{\rm D}}},
\end{equation}
where $\bar Y$, is the experimental charge normalized yield, and $\bar Y_{\rm MC}$ is the charge normalized Monte Carlo equivalent yield, and the denominator is the ratio of the yields from the $^2$H target. To reduce the impact of non-isoscalar effects such as $K^+\Sigma^-$ production off a neutron, we consider it more appropriate to define transparency as the ratio of nuclear cross sections with respect to a deuterium target instead of a free proton target. Since the Monte Carlo simulation does not include final-state interactions between the kaon and the residual nucleons, the nuclear transparency is a measure of the absorptive effect of such final-state interactions. The kaon transparencies are listed in Table~\ref{table2}. For completeness, in Table~\ref{table2} we have also listed the transparencies extracted as ratio of nuclear cross sections with respect to $^1$H target. 

\begin{figure}[h]
	\centering
	\includegraphics[width=86mm]{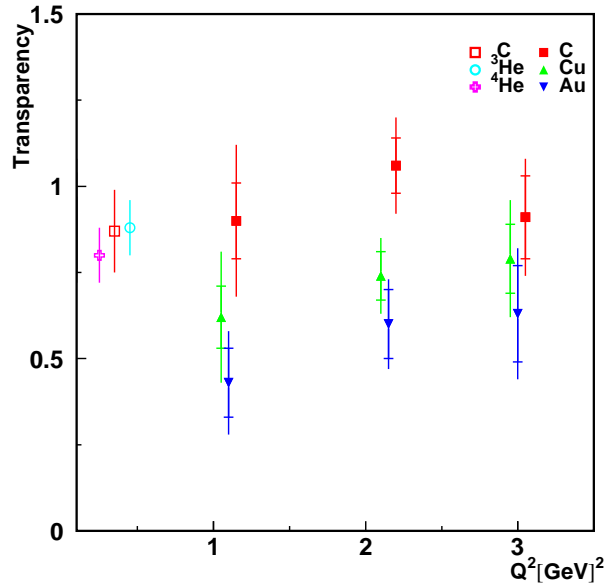}
	\vspace{0.43ex}
	\caption{(Color online) Nuclear transparency for kaons (as defined in the text) for different targets vs. $Q^2$. The Copper and Carbon points are shifted by -0.05 and 0.05 GeV$^2$ in $Q^2$, respectively, for ease of display. The inner error bars are the statistical uncertainties only, the outer error bars are the sum in quadrature of statistical and systematic uncertainties. The solid  points are the results from this experiment. Results from a previous experiment on light nuclei~\cite{eek1,eek2} at $Q^2$=0.35 GeV$^2$, modified to display equivalent quantities (see text), are shown as open symbols.}
%Results from a previous experiment on light nuclei~\cite{eek1,eek2} at $Q^2$ = 0.35 GeV$^2$ are shown as open symbols.}
	\label{fig:transp2_itr1}
\end{figure}

The transparencies and their $Q^2$ dependence are shown for the three different heavy target nuclei in Fig.~\ref{fig:transp2_itr1}. In an earlier experiment at JLab the quasi-free electroproduction of kaons from light nuclei was studied at $Q^2 =$ 0.35 GeV$^2$, and an effective proton number was extracted from these data~\cite{eek2}. The effective proton number was defined as the ratio  of kaon electroproduction cross section on light nuclei to the cross section on hydrogen. In Fig.~\ref{fig:transp2_itr1} we have shown these effective proton numbers along with our results. In order to display equivalent quantities, we have shown the ratio of the effective proton number from light nuclei to that from $^2H$. The results from the two experiments are consistent with each other. %In order to display equivalent quantities we have taken the ratio of the proton number from light nuclei to that from $^2H$ as obtained from~\cite{eek2}. The results from these two experiments are consistent. 
Within the experimental uncertainties the nuclear transparency of kaons do not exhibit any energy dependence. To guide further interpretation we have analyzed the nuclear and energy dependence of these transparency data in terms of two simple one-parameter descriptions as detailed below.

\begin{tiny}
\begin{table*}
 \renewcommand{\arraystretch}{1.2}
 \centering
\caption{(left) The ratio of the nuclear cross sections with respect to the $^1$H cross sections at three different values of $Q^2$; (right) the kaon transparency, or
the ratio of (heavy) nuclear cross sections with respect to the $^2$H cross sections. The ratios are shown with both their statistical and systematic uncertainties which include model uncertainties.}
  \begin{tabular}{|c|c|c|c|c||c|c|c|}\hline
   & \multicolumn{4}{|c||}{Ratio of nuclear and $^1$H cross sections}& \multicolumn{3}{|c|}{Ratio of nuclear and $^2$H cross sections}\\ \hline
 $Q^2$ & $^2$H & Carbon & Copper & Gold & Carbon & Copper & Gold\\\hline
   1.1 &0.88$\pm$0.09$\pm$0.11 & 0.80$\pm$0.07$\pm$0.13 & 0.55$\pm$0.04$\pm$0.07 & 0.39$\pm$0.04$\pm$0.03 &0.91$\pm$0.10$\pm$0.17 & 0.62$\pm$0.06$\pm$0.10 & 0.43$\pm$0.04$\pm$0.05\\\hline
   2.2 & 0.86$\pm$0.06$\pm$0.06 & 0.91$\pm$0.07$\pm$0.09 & 0.64$\pm$0.04$\pm$0.04 & 0.51$\pm$0.05$\pm$0.03 & 1.06$\pm$0.09$\pm$0.11 & 0.74$\pm$0.05$\pm$0.06 & 0.60$\pm$0.06$\pm$0.05\\\hline
	 3.0 & 0.92$\pm$0.09$\pm$0.11  & 0.84$\pm$0.10$\pm$0.07  & 0.73$\pm$0.07$\pm$0.07  & 0.58$\pm$0.08$\pm$0.05 & 0.91$\pm$0.11$\pm$0.12  & 0.79$\pm$0.08$\pm$0.11  & 0.63$\pm$0.09$\pm$0.08\\\hline
\end{tabular}
\label{table2} 
\end{table*}
\end{tiny}

%mysection
We describe the measured transparencies in terms of the simple geometrical model outlined in Ref.~\cite{jlabp2}. This model assumes classical
attenuation of hadrons propagating in the nucleus, with an effective hadron-nucleon cross section $\sigma_{eff}$ that is {\sl independent} of density,
and the transparency can be written as:
\begin{equation} \label{effcrr}
T_{hadron} = 
{\frac{1}{A}}
{\int d^{3}r \rho_{A}(\vec{r})exp{\left[-\int dz'\sigma_{eff}\rho_{A-1}(\vec{r}^\prime)\right]}},
\end{equation}
with $\rho_A$ representing the nucleon number density in the nucleus $A$. The nucleon is struck at position $\vec{r} =(x,y,z)$ and the direction of the outgoing hadron is labeled $z^{'}$, with $\rho_{A-1}(\vec{r}^{'})$ as the nucleon number density of the recoil $A-1$ system at the position $\vec{r}^{'}=(x,y,z^{'})$ on the hadron's path. Regardless of its limitations due to the simplicity of the geometrical model,
we hope to gain insight in the relative behavior of the effective cross sections versus both momentum transfer and other electro-produced hadrons.

\begin{figure*}
  \centering
%  \begin{center}
  \includegraphics[width=178mm]{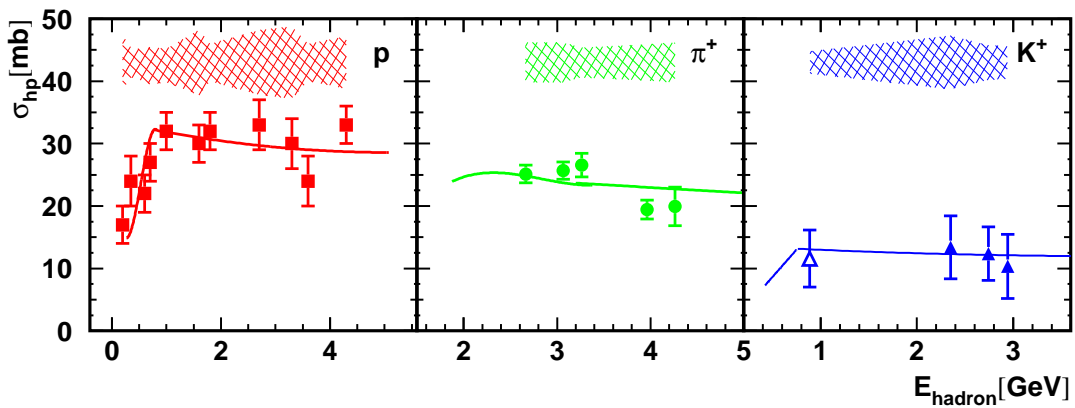}
  \vspace{-30ex}
  \caption{(Color online) Effective cross sections as a function of the hadron kinetic energy for protons, pions and kaons. The solid  points are the results from this experiment. Results from a previous experiment on light nuclei~\cite{eek1,eek2} at $Q^2$ = 0.35 GeV$^2$ are shown as open symbols.
The solid curves are fits to the global proton-nucleon (red), pion-nucleon (green) and kaon-nucleon scattering data (blue)~\cite{pdg},
scaled by 0.68~(0.03), 0.81~(0.03) and 0.65~(0.14), respectively. The numbers within parentheses are the uncertainties associated with the scale factors. The data shown are the average over different targets, weighted by the statistical uncertainty only, and the standard deviation shown by the hatched bands.}
  \label{fig:crr2} 
%  \end{center}
\end{figure*}

We proceeded to extract such effective cross section by fitting the measured transparencies to the model described. For this, we took the nuclear (charge)
density distributions from Ref.~\cite{rhodist}. The effective cross section is then the only free parameter. Finally, we take the average of the effective cross sections obtained for $^{12}$C, $^{56}$Fe and $^{197}$Au nuclei, weighted by their respective statistical uncertainty. The standard deviation for each hadron is shown by the hatched bands in Fig.~\ref{fig:crr2}.
%Only the statistical uncertainties of the transparencies are used when extracting the effective cross sections.
In addition to the kaon, we have shown the effective cross section for protons obtained from Ref.~\cite{jlabp2} and we have similarly calculated the effective cross sections from the published pion transparencies~\cite{jlabpi2}. 
%In order to be consistent with the kaon analysis the pion transparencies for $A>2$ targets were divided by the respective transparencies for deuterium. 
All results are shown as a function of the hadron kinetic energies in Fig.~\ref{fig:crr2}. For comparison, we also show a fit (solid curves) to the experimental free hadron-nucleon cross section data from Ref.~\cite{pdg}. The effective cross sections extracted from the proton transparencies are in excellent agreement, in their dependence on kinetic energy, with the free proton-nucleon case. The absolute magnitude is under-predicted, but good agreement can be obtained by scaling the free proton-nucleon cross section by 0.68~(0.03). To a lesser extent, the kinetic energy dependence of the effective pion-nucleon cross sections
extracted from pion transparency data also agrees with that expected from free pion-nucleon scattering. For pions the scale factor was 0.81~(0.03). The free kaon-nucleon cross sections were scaled by 0.65~(0.14) to agree with the extracted effective kaon-nucleon cross sections. The numbers shown within parentheses are 
the uncertainty associated with the scale factors. 

\begin{figure*}
\begin{center}
\includegraphics[width=178mm]{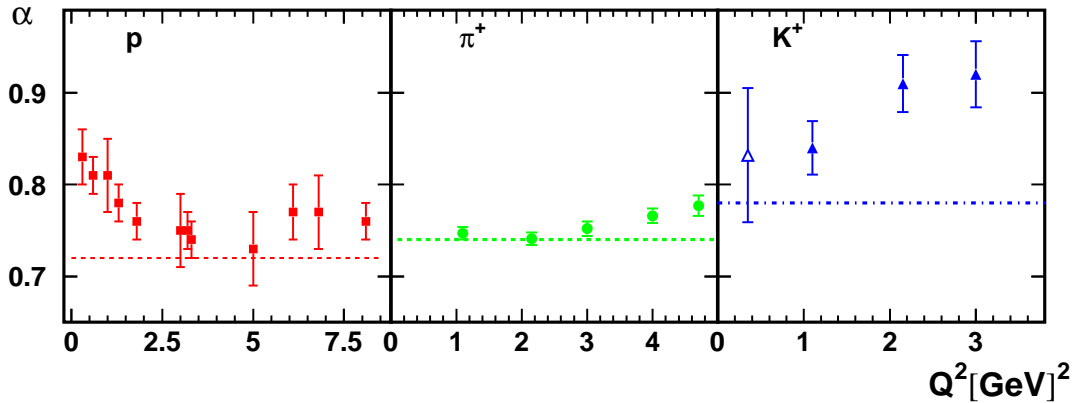}
\vspace{-30ex}
\caption{(Color online) The parameter $\alpha$ (see text) as a function of $Q^2$ for protons (squares), pions (circles) and kaons (triangles) obtained from electron
scattering. The solid  points are the results from this experiment. Results from a previous experiment on light nuclei~\cite{eek1,eek2} at $Q^2$ = 0.35 GeV$^2$ are shown as open symbols. The dashed lines shows the value of $\alpha$ for protons, pions and kaons respectively obtained from hadron-nucleus scattering~\cite{carroll}.}
\label{fig:carolplot}
\end{center}
\end{figure*}

It is interesting to note that the energy dependence of the effective cross section for the protons and pions is consistent with the energy dependence of the
free cross sections (for kaons the large experimental uncertainties are too large to make any strong conclusions), however the absolute magnitude of the effective cross section is reduced. This renormalization is interpreted~\cite{ne18} to take into account effects that can not be absorbed in the density-independent assumptions of the simple geometric model. For instance, the simple model does not account
for effects such as Pauli blocking and short range correlations. In general, one would expect these effects to be the smallest for kaons since they introduce
a strangeness degree of freedom~\cite{waas97}. However, in contrast we find that the kaon-nucleon effective cross sections extracted from electroproduction is significantly smaller than the free cross section, with the reduction larger than that for pions and comparable to the reduction for protons. We note that, for both protons and pions, more sophisticated models~\cite{fsz94,larson06} have been shown to be in good agreement with the absolute magnitude of the measured transparencies, for example see Ref.~\cite{jlabp2,jlabpi1} for more details. Such theoretical calculations are currently unavailable for kaons.

In addition to the $Q^2$ dependence, we examined the dependence of the measured kaon transparencies on the atomic mass number $A$ {\sl at fixed $Q^2$} with
a single-parameter fit form, $T = (A/2)^{\alpha-1}$. Here $\alpha$ is used to parametrize the nuclear cross section. This simple ansatz is equivalent to the more commonly used $\sigma_N = \sigma_0A^{\alpha}$ parameterization, with $\sigma_N$ the nuclear cross section and $\sigma_{0}$ the elementary (isoscalar) nucleon cross section~\cite{carroll}. The extracted values of the parameter $\alpha$ are shown in Fig.~\ref{fig:carolplot}. 
We have included the $\alpha$ values from fits to the proton and pion transparency data~\cite{jlabp2,jlabpi2}. In order to be consistent with the kaons, the proton and pion transparencies for $A>2$ targets were recalculated as the ratios of cross sections from nuclear targets to deuterium and the recalculated transparencies were fit to the single parameter form, $T = (A/2)^{\alpha-1}$. This recalculation will render values for $\alpha$ 
slightly different from our previous publications for protons and pions. In addition we show, as lines, the $\alpha$-values as extracted from high-energy hadron-nucleus collisions by Carroll {\it et al.}~\cite{carroll}, for kaons, pions and protons. For all three hadrons, we find the $\alpha$ value as extracted from the electron scattering experiments to be somewhat larger than those from hadron-nucleus collisions. This can be partly attributed to the nature of the probe used in these collisions: the strong hadronic probe is more surface-dominated, whereas the electromagnetic probe likely samples the entire nuclear volume.  For kaons, the parameter $\alpha$ from electro-production is significantly larger than the value extracted from high-energy kaon-nucleus scattering. This is contrary to the traditional nuclear physics expectations, however, the relatively large experimental uncertainties and the lack of corresponding energy dependence in the nuclear transparency do not allow strong conclusions regarding Quantum Chormodynamic effects such as the formation of compact states.

%%%%%%%%%%%%%%%%%%%%%%%%%%%%%%%%%%%%%%%%%%%%%%%%%%%%%%%%%%%%%%%%%%%%%%%%%%%%%%%%%%%%%%%%%%%%%%%%%%%%%%%
%%%%%%%%%%%%%%%%%%%%%%%%%%%%%%%%%%%%%%Summary and Acknowledgments%%%%%%%%%%%%%%%%%%%%%%%%%%%%%%%%%%%%%%%
%%%%%%%%%%%%%%%%%%%%%%%%%%%%%%%%%%%%%%%%%%%%%%%%%%%%%%%%%%%%%%%%%%%%%%%%%%%%%%%%%%%%%%%%%%%%%%%%%%%%%%%%
In summary, we measured the reaction $A(e, e'K^+)$ for $^1$H, $^2$H, $^{12}$C, $^{63}$Cu and $^{197}$Au at $Q^2$ = 1.1, 2.2 and 3.0 GeV$^2$.
We extracted the nuclear transparency of kaons as the ratio of the kaon electro-production cross sections of the heavy nuclei with deuterium.
The energy dependence of the nuclear transparency is found to be consistent with traditional nuclear physics expectations within experimental uncertainties.
However, the absolute magnitude of the effective kaon-nucleon cross sections extracted from the nuclear transparency are found to be substantially smaller
than the free kaon-nucleon cross sections in a simple geometric model. Effective hadron-nucleon cross sections for protons, pions and kaons analyzed in
similar manner find reduction factors of 0.68(0.03), 0.81(0.03) and 0.65(0.14), respectively, as compared to the free hadron-nucleon cross sections. The kaon results are
significantly smaller than the effective kaon-nucleon cross sections as obtained from kaon-nucleus scattering. The $A$ dependence of the kaon-nucleon
cross section extracted from the transparency measurements were also found to be significantly different from those obtained from kaon-nucleus scattering.
The difference between kaon electroproduction and kaon-nucleus scattering results are significantly larger than the earlier differences observed for pions
and protons. Due to the smaller absolute values of the free kaon-nucleon cross sections (as compared to the pion and proton), the opposite was expected.
This could be an indication that these results cannot all be simultaneously explained in terms of traditional nuclear physics effects, however, the relatively large experimental uncertainties and the lack of energy dependence in the nuclear transparency do not allow strong conclusions.
It would be of great interest to extend a recent calculation of kaon production data in heavy-ion collisions, going beyond the conventional meson-nucleon
nuclear many-body calculations in the framework of the quark-meson coupling model~\cite{sibirtsev}, also to these results.

We acknowledge the outstanding support of JLab Hall C technical staff and Accelerator Division in accomplishing this experiment. This work was supported
in part by the U.~S.~Department of Energy, the U.~S.~National Science Foundation, and the Natural Science and Engineering Research Council of Canada.
This work was supported by DOE contract DE-AC05-84ER40150 under which the Southeastern Universities Research Association (SURA) operates the Thomas Jefferson
National Accelerator Facility.
%\end{acknowledgments}

%\bibliography{paper}

\end{document}